\documentclass[pra,amsmath,amssymb,twocolumn,superscriptaddress]{revtex4}

\usepackage{amsmath,amssymb}
\usepackage[usenames]{color}
\usepackage[colorlinks,linkcolor=blue,citecolor=blue]{hyperref}
\usepackage[T1]{fontenc}
\usepackage{grffile}
\usepackage[pdftex]{graphicx}
\usepackage{xspace}
\usepackage{enumitem}
\usepackage{xcolor} 
\usepackage{bm}       

\newcommand{\subscript}[2]{$#1 _ #2$}
\newcommand{\uw}{\uparrow}
\newcommand{\dw}{\downarrow}

\newcommand{\proj}[1]{| #1 \rangle\langle #1 | }
\newcommand{\ave}[1]{\langle #1 \rangle }

\newcommand{\tr}{{\rm tr}}
\newcommand{\im}{{\rm i}}

\newcommand{\A}{\mathcal{A}}
\newcommand{\B}{\mathcal{B}}
\newcommand{\C}{\mathcal{C}}
\newcommand{\D}{\mathcal{D}}

\definecolor{dbc}{rgb}{0.,0.4,.8} 
\definecolor{ogc}{rgb}{0.4,.6,.5} 

\newcommand{\sutd}{Science and Math cluster, EPD Pillar, Singapore University of Technology and Design, 8 Somapah Road, 487372 Singapore}
\newcommand{\cqt}{Centre for Quantum Technologies, National University of Singapore, Singapore 117543}
\newcommand{\nus}{Department of Physics, National University of Singapore, Singapore 117546}
\newcommand{\nusg}{NUS Graduate School for Integrative Science and Engineering, Singapore 117597}
\newcommand{\augsburg}{Institute of Physics, University of Augsburg, Universitatstra{\ss}e 1, D-86135 Augsburg, Germany}
\newcommand{\munich}{Nanosystems Initiative Munich, Schellingstra{\ss}e 4, 80799 Munchen, Germany}

\begin{document}

\title{Single atom energy-conversion device with a quantum load}

\author{Noah Van Horne}%
\affiliation{\cqt}%
\author{Dahyun Yum}%
\affiliation{\cqt}%
\author{Tarun Dutta}%
\affiliation{\cqt}%
\author{Peter H\"anggi}%
\affiliation{\cqt}
\affiliation{\augsburg}%
\affiliation{\nus}
\affiliation{\munich}
\author{Jiangbin Gong}\email{phygj@nus.edu.sg}
\affiliation{\nus}%
\affiliation{\nusg}%
\author{Dario Poletti}\email{dario_poletti@sutd.edu.sg}
\affiliation{\sutd}%
\author{Manas Mukherjee}\email{phymukhe@nus.edu.sg}
\affiliation{\cqt}%
\affiliation{\nus}%

\date{\today}

\begin{abstract}
This work reports on a single atom energy-conversion device, operating either as a quantum engine or a refrigerator, coupled to a quantum load. The ``working fluid'' is comprised of two optical levels of a single ion, and the load is one vibrational mode of the same ion which is cooled down to the quantum regime. The energy scales of the optical and vibrational modes differ by 9 orders of magnitude. 
We realize cyclic energy transfers between the working fluid and the quantum load, either increasing or decreasing the population of the vibrational mode. This is achieved despite the interaction between the load and the working fluid leads to a significant population redistribution and quantum correlations between them. The performance of the device is examined as a function of several parameters, and found to be in agreement with theory. We specifically look at the ergotropy of the load, which indicates the amount of energy stored in the load that can be extracted with a unitary process. We show that ergotropy rises with the number of engine cycles despite an increase in the entropy of the load. Our experiment represents the first fully quantum $4-$stroke energy-conversion device operating with a generic coupling to a quantum load.
\end{abstract}%

\maketitle

\paragraph*{Introduction:}  Our green future may rely on energy conversion devices at such scales and temperatures that quantum effects become relevant or even dominant. Nanoscale heat engines and refrigerators are hence recognized as a promising research frontier \cite{benenti2017fundamental, Kosloff2012, GelbwaserKurizki2015, GooldSkrzypczyk2016}. 
To pave the way towards technological breakthroughs, researchers have investigated a number of fundamental questions, even examining the validity and pertinence of thermodynamic concepts at the nanoscale. Indeed, recent years have seen fruitful studies exploring how the discreteness of energy levels, quantum statistics, quantum adiabaticity, measurement, coherence and entanglement affect the operation of heat engine cycles~\cite{vuletic2018, zheng2015quantum, gong2014interference, xiaogong2015, CottetHuard2017, yi2017single, PhysRevE.98.042122, watanabe2017quantum, RouletScarani2017, uzdin2015equivalence}. Models of quantum heat engines have also been exploited as a platform to investigate the thermodynamics underlying nanoscale energy conversion beyond the weak coupling limit~\cite{Campisi, HanggiReview, HanggiReviewErratum, Jarzynski2, Segal, TalknerHanggi2016, Jarzynski}, or in the presence of engine-load correlations \cite{scully2001extracting, teo2017converting}. Experimental investigations of quantum engine cycles are hence called for to clearly assess the correspondences, or lack thereof, between classical and quantum thermodynamic engine cycles. This experimental work aims to set an important milestone towards the realization and understanding of single atom energy-conversion cycles acting on a quantum load. To the best of our knowledge, our experimental system is the first realization and thermodynamic analysis of a fully quantum engine and refrigerator cycle simulator with generic coupling to a quantum load. 

Quite recently, several experimental realizations of nanoscale engine cycles and thermodynamic processes have been reported \cite{blickle2012realization, rossnagel2016single, ronzani2018tunable, maslennikov2017quantum, klatzow2017experimental, peterson2018experimental, von2018spin}, including experiments in the quantum regime \cite{maslennikov2017quantum, klatzow2017experimental, peterson2018experimental, von2018spin}.
In this work we build upon previous efforts in this direction by realizing an important, hitherto unimplemented class of heat engine cycles. In particular, the working fluid of our engine cycle consists of two optical states of a single trapped ion, and the load is one vibrational mode of the same ion. The load is prepared with about only one phonon on average, hence operating in a quantum regime. Importantly, the coupling of the engine to the load is via a generic interaction which does not commute either with the engine or the load, thus enabling a significant back action between the two. The genereic nature of the interaction allows us to investigate the impact of the strong engine-load coupling on the energy flow into the load and on the dynamics of the entropy of the load. 
Furthermore, the engine cycle can be readily executed in reverse, thus functioning as a refrigerator which can decrease both the energy and, because of the coupling, the entropy of the load.          

\begin{figure}[h!]
\begin{center}
\includegraphics[width=1\columnwidth]{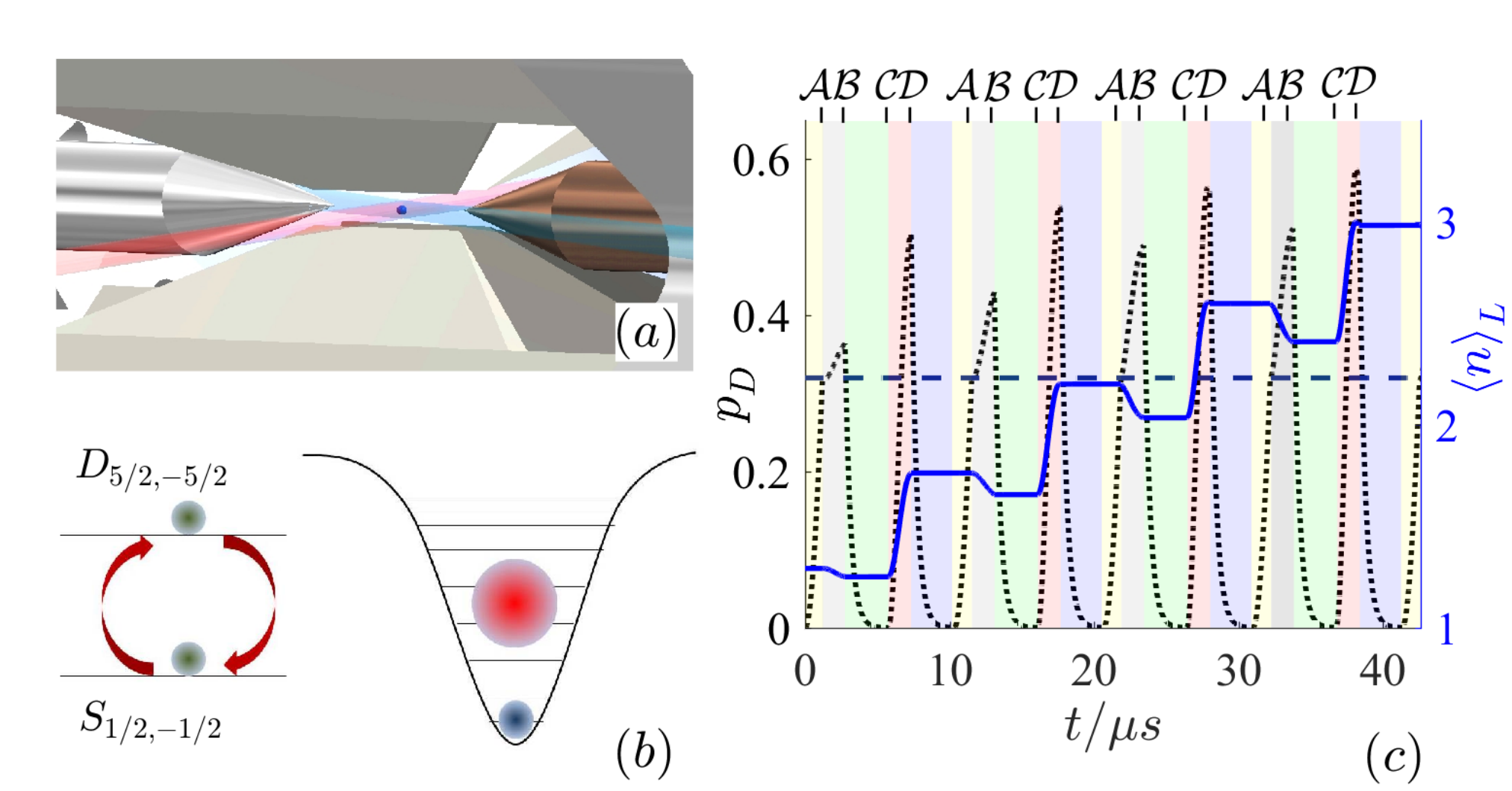}
\caption{\label{fig:fig1} (a) Image of the ion trap set-up with a single ion acting as both an engine and a load, coupled together by laser beams (red and blue lines). (b) Schematic representation of the working fluid of a quantum engine (two-level system, left) and the load (the vibrational modes of a quantum harmonic oscillator, right). The two-level system undergoes a cyclic evolution. The circles in the harmonic oscillator represent the occupation probabilities of different modes. Initially, when the ion is cooled near the ground state, the spread in occupation probabilities is small (smaller blue circle). After operating the energy-conversion device for a number of forwards (engine-type, as opposed to refrigerator-type) cycles, the spread in occupation probabilities increases (larger, fuzzier, red circle). (c) The black dotted line shows realistic numerical simulations for the evolution over time of the occupation probability of the higher energy level ($D_{5/2,-5/2}$), of the two-level system, referred to as $p_D$ (left vertical axis). The solid blue line shows the evolution of the mean phonon number $\langle n \rangle_L$ (right vertical axis). The four strokes of the cycle are highlighted by shadings of different colors. Stroke (I), from state $\D$ to $\A$, is divided into two portions, in blue for (I$_{a}$), and yellow for (I$_{b}$). Stroke (II), from $\A$ to $\B$ is in grey, stroke (III), from $\B$ to $\C$ is in green, and stroke (IV) from $\C$ to $\D$ is pink. The horizontal dashed line in the middle shows the value to which the engine population is reset, once in each cycle, at the end of stroke (I).  For Fig.~1(c) this value is $p_D^\A=0.32$.}
\end{center}
\end{figure}

Fig.~\ref{fig:fig1}(a) depicts our experimental platform consisting of a linear ion trap, along with two addressing lasers. Fig.~\ref{fig:fig1}(b) (left side) presents a schematic of the two-level working fluid (in the following also referred to as the {\it engine}), which is coupled to one of the ion's vibrational modes, which is the {\it load} (Fig.~\ref{fig:fig1}(b), right side). 
Cycles of the quantum engine are implemented using lasers, which either reset the engine or couple the engine to the load. While the structure of our energy-conversion device is inspired by the classical Otto cycle, there exist three fundamental differences between our engine cycle and typical classical thermodynamic cycles. First, throughout one cycle the engine and the load build up quantum correlations (discussed in appendix F). This results in decoherence of, individually, the engine and the load. It also leads to entropy generation, accompanied by a transfer of entropy between the engine and the load. Second, the back action of the load causes significant population redistribution within the engine. This is common to nanoscale/quantum engines working against a quantum load. Third, through a process involving both work (from the lasers) and heat exchange (with the vacuum photon bath), during the resetting steps of each cycle, the engine state is brought back to one of two predetermined nonequilibrium states, rather than a thermal equilibrium state. The three aforementioned features mean that the cycle implemented by our device cannot be mapped directly to typical classical thermodynamic engine cycles. This allows us to explore unprecedented situations relevant to actual energy-conversion devices in the quantum regime. 

Such an engine cycle cannot be mapped directly to typical classical thermodynamic engine cycles, and allows us to explore unprecedented situations relevant to actual energy-conversion devices in the quantum regime.
Furthermore, our design makes it possible to realize cyclic energy transfer between a working fluid (optical states) and a load (vibrational mode) despite energy scales differing by $9$ orders of magnitude.

\paragraph*{The setup:} The device consists of a single Ba$^+$ ion confined in a radiofrequency linear Paul trap \cite{yum2017optical}. The Hamiltonian of the whole system is given by  
\begin{align}
H(t)=H_E + H_L + V(t),
\end{align}
where $H_E=\hbar(\nu/2)\sigma^z$ is the Hamiltonian describing the two optical levels of the engine, $H_L=\hbar\omega (n+1/2)$ is the Hamiltonian of the load and $V(t)$ describes the engine-load coupling, which can be of Jaynes-Cummings (JC) type, $V(t)=\hbar g(t)(\sigma^+a + \sigma^-a^{\dagger})$, or of anti-Jaynes-Cummings (AJC)  type $V(t)=\hbar g(t)(\sigma^-a + \sigma^+a^{\dagger})$. Here $\hbar \nu$ denotes the energy difference between the two levels of the engine, $\sigma^z$ is the standard Pauli matrix, $\omega$ is the harmonic oscillator frequency of the load and $n$ is the phonon number operator for the load. Additionally, $g(t)$ denotes the time-dependent coupling strength, $\sigma^+$ ($\sigma^-$) raises (lowers) the engine state by one photon, and $a$ ($a^\dagger$) destroys (creates) one phonon in the load. Note that $V(t)$ does not commute with either the engine Hamiltonian $H_E$ or the load Hamiltonian $H_L$, which indicates that the engine-load coupling is capable of redistributing the population between the engine states, i.e. the two optical levels
 $S_{\frac{1}{2},-\frac{1}{2}}$ and $D_{\frac{5}{2},-\frac{5}{2}}$ (see Appendix A for details). 

\paragraph*{The engine cycle:}
We realize a $4-$stroke cycle. Analogously to an Otto cycle, in two of the four strokes the engine undergoes energy exchange with the laser field as well as the environment (dephasing, heating and spontaneous emission as detailed in Appendix D), while in the absence of engine-load coupling. In the other two strokes, the engine-load coupling is switched on.
Specifically, each cycle executes the following four steps:
(I) With the engine decoupled from the load, the engine is reset to a fixed quantum state consisting of a pure superposition of the two optical levels. (II) The engine-load coupling of the JC type is switched on, and the quantum number of the load is either increased or decreased by one, depending on whether the engine state is in the lower or upper level. This step also creates entanglement between the load and the engine.  (III) The engine and the load are decoupled, and the component of the engine which is in the excited state is dissipatively brought back to the ground state. (IV) The engine-load coupling of the AJC type is switched on.
Stroke (I) brings the engine back to state $\A$ (Fig.~\ref{fig:fig1}(c)), and then each of the subsequent strokes brings the engine to, in consecutive order, states $\B$, $\C$ and $\D$ (details available in Appendices B and C). If not explicitly stated otherwise, the specific time-dependent driving protocol used to implement strokes (II) and (IV) is a resonant sideband transfer \cite{LeibfriedWineland2003} (see Appendix B). It should be highlighted that if we operate the cycle in the reverse order (discussed below), the above-described engine cycle can be regarded as the first experimental realization of a cyclic quantum refrigerator operating in the strong engine-load coupling regime.      

\begin{figure}[h!]
\begin{center}
\includegraphics[width=1\columnwidth]{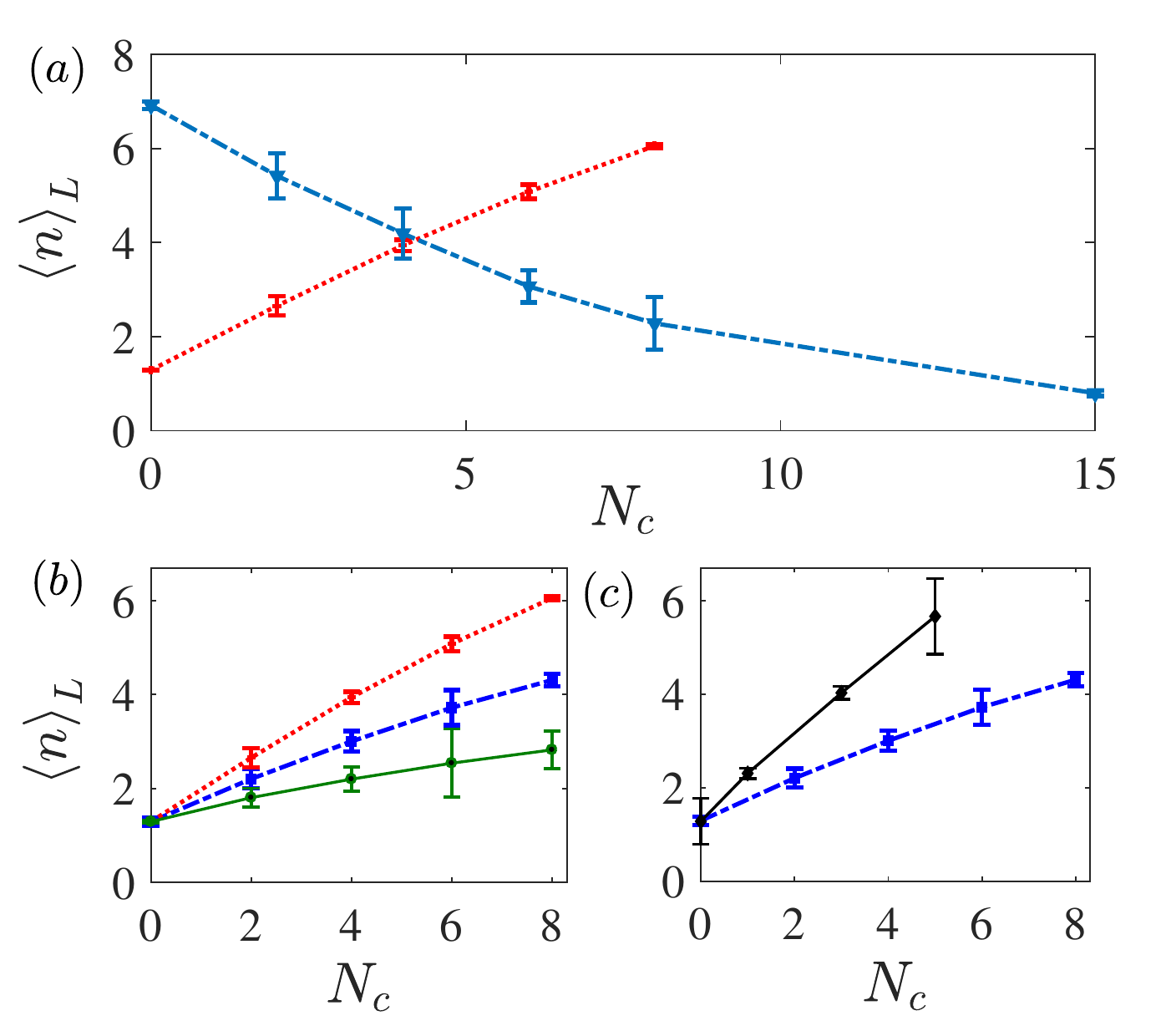}
\caption{\label{fig:fig2} (a-c) Experimentally measured mean phonon occupation number of the load $\langle n \rangle_L$ versus the number of engine cycles $N_c$.
Panel (a) compares an engine cycle (red dotted line with {\color{red}$\bm{+}$}) with a reversed cooling cycle (blue dot-dashed line with {\color{dbc}$\blacktriangledown$}), with $p_D^\A=0.5$ in both situations.
(b) Performance of engine cycles characterized by different $p_D^\A$, i.e. the population in state $D$ upon state resetting in stroke (I), with $p_D^\A=0.5$ (red dotted line with {\color{red}$\bm{+}$}), $p_D^\A=0.32$ (blue dot-dashed line with {\color{blue}$\blacksquare$}) and $p_D^\A=0.15$ (green continuous line with {\color{ogc}$\bullet$}). For larger $p_D^\A$, the rate of increase of $\langle n \rangle_L$ per engine cycle is greater. (c) Performance of engine cycles for two different engine-load coupling protocols, namely, rapid adiabatic passage (black $\blacklozenge$) or resonant sideband transfer (blue {\color{blue}$\blacksquare$}), and with $p_D^{\A}=0.32$.
In all panels, the lines connecting the data points are only to guide the eye. The error bars contain both statistical and systematic errors, obtained by parameter estimation using the least-squares method (see Appendix E for details).  }
\end{center}
\end{figure}

The above four-stroke cycle deserves a few remarks. 
In stages (II) and (IV) the coupling between the engine and the load builds up correlations, as well as quantum entanglement, between them (see Appendix F). 
Hence, from the perspective of the engine itself, the strokes in which the engine is coupled to the load are not described by a unitary evolution. This is in sharp contrast to many existing theoretical models of quantum engine cycles where an operation to extract work from the engine is typically assumed to be a unitary process (e.g.\cite{zheng2015quantum, xiaogong2015, yi2017single, watanabe2017quantum}).
Hence it is of crucial importance to include a quantum load in the picture in order to realistically assess the performance of nanoscale engine cycles. 
Next we turn to stroke (I), where the engine state is reset. Thus, during the optical pumping, a dissipative process, heat is transferred from the engine to the environment. During the coherent laser excitation, a unitary process  work is done on the engine by the lasers. Thus, there are both heat and work transfers with the engine during this stroke. 
The state upon resetting can be a pure state, when fixing the laser phase used for coherent excitation (which is what is done in the experiment). Alternatively, it can effectively be a mixed state when randomizing the same laser phase. 
We note that it is not possible to define a temperature for the engine since in general it is not in thermal equilibrium with the environment (see Appendix D). 
Lastly, we mention how the load is prepared in its initial state. This is done by successively applying Doppler cooling and motional sideband cooling. The ion is brought to a thermal state with average phonon occupation number $\ave{n}_L \approx 1.2$ which confirms the quantum nature of the load. Note that we use the notations $\ave{\dots}_L=\tr(\rho^{}_L \dots)$ where $\rho_L^{}=\tr_E(\rho_{_{E+L}})$ is the reduced density matrix of the load obtained by tracing out the engine from the engine$+$load density matrix $\rho_{_{E+L}}$.    

\paragraph*{Results:} We first analyze the system using numerical simulations, taking into account the most relevant aspects of our experimental set-up, including the main decoherence effects (see  Appendix D for more details). The black dotted line in Fig.~\ref{fig:fig1}(c) shows simulation results for the occupation of the $D$ level, $p_D$, over time, while the solid blue line shows the average phonon number in the load, $\ave{n}_L$. Different shadings identify different strokes, with the grey, light green, and pink regions corresponding respectively to strokes (II), (III), and (IV). Stroke (I) is partitioned into two shadings: blue for resetting the engine to its ground state (I$_a$), and yellow for bringing the engine to a predetermined value of its excited-state population, $p_D^\A$, (I$_b$). 
The horizontal black dotted line in Fig.~\ref{fig:fig1}(c) marks the value corresponding to the end of the yellow region, which for this set of simulations is $p_D^\A = 0.32$. In Fig.~\ref{fig:fig1}(c), $\ave{n}_L$ increases roughly proportionally to the number of engine cycles. Meanwhile, the occupation of the D-level of the engine is seen to settle down on a periodic pattern after a transient period.  

Next we report the main experimental results. Fig.~\ref{fig:fig2}(a) shows the average phonon number, $\langle n \rangle_L$, versus the engine cycle number $N_c$ (red dotted line with $+$),  confirming that our engine cycle is indeed capable of transferring energy to the load in an approximately linear fashion, with the average phonon occupation number increasing from $\langle n \rangle_L=1.2$ to $\langle n \rangle_L=6.1$ over the course of $N_c=8$ cycles. To demonstrate that our engine cycle can operate in reverse, we execute the strokes in the order (I), (IV), (III), (II) and then repeat the cycle, with the load initially prepared in a highly excited state. The results for this refrigeration operation (blue triangles) are also presented in Fig.~\ref{fig:fig2}(a) for comparison. In both cases, the change of $\langle n \rangle_L$ is fairly uniform from cycle to cycle for the first $8$ cycles. 

To measure the average phonon number we used the experimentally-derived probability distribution of the occupation of each level of the load, $p_n$. This is obtained by least-squared fitting of $p_n$ with experimentally measured blue sideband Rabi oscillation, and using the numerically derived distribution of $p_n$ as initial condition for the fitting (details in Appendix E). Each data point is based on $150$ repetitions of the experiment, averaged together.

We now focus on the performance of the forward cycle. In Fig.~\ref{fig:fig2}(b), we show that the rate of change in $\langle n \rangle_L$ can be tuned by resetting the engine to different states before stroke (II) (point $\A$ in Fig.~\ref{fig:fig1}(c)), or in other words by varying $p^{\A}_D$. An increase in $p^{\A}_D$ increases the rate of change in $\langle n \rangle_L$, and equates to greater energy flow into the load. This is because for larger $p^{\A}_D$, the JC-type engine-load coupling in stroke (II) induces a smaller decrease of $\ave{n}_L$.

Finally, we look at how different engine-load coupling protocols might affect the energy flow from the engine to the load. The experimental results for this comparison are shown in Fig.~\ref{fig:fig2}(c), where, in terms of $\langle n \rangle_L$ {\it vs} $N_c$, a rapid adiabatic passage (RAP) protocol \cite{WatanabeUrabe2004} (black $\diamond$) is compared against the resonant sideband transfer (blue $\square$) used in Fig.~\ref{fig:fig2}(a-c). For both cases in Fig.~\ref{fig:fig2}(c), $p_D^{\A}=0.32$. Remarkably, compared with the resonant side band transfer protocol, the RAP protocol nearly doubles the energy transfer rate. This observation can be understood by the fact that the RAP protocol effectiveness in transferring the populations is much less sensitive to the load distribution $p_n$ compared to the resonant sideband transfer.

\begin{figure}[h!]
\begin{center}
\includegraphics[width=\columnwidth]{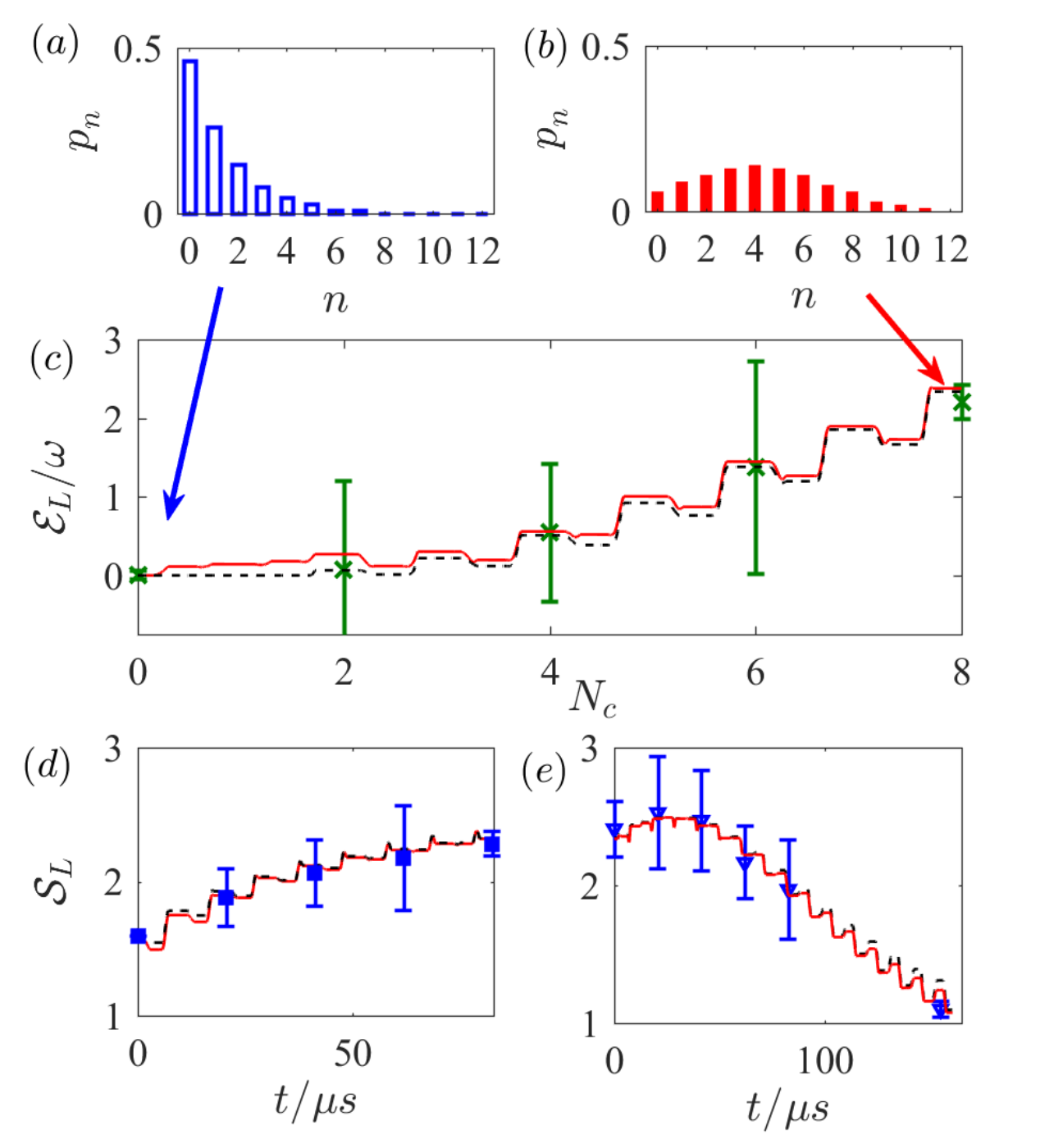}
\caption{\label{fig:fig3} (a,b) Experimentally derived occupation probabilities $p_n$ of the load, upon initial preparation and after $N_c=8$ cycles, for the stroke (I) resetting parameter $p_D^\mathcal{A}=0.5$. (c) Ergotropy of the load $\mathcal{E}_L$ versus $N_c$, with $p_D^\mathcal{A}=0.5$. The red continuous line shows the exact numerical value. The black dashed line is the approximate ergotropy based on only the diagonal elements of the reduced density matrix of the load.  The green $\times$ are the experimental data. (d) Entropy of the load $\mathcal{S}_L$ versus time, for the exact numerical estimate (red continuous line), an approximate numerical estimate considering only the diagonal elements of the density matrix (black dashed line), and the experimentally evaluated entropy values after 2, 4, 6, and 8 cycles (blue squares), for $p_D^\mathcal{A}=0.32$. (e) The same as (d), but for the refrigerator cycle, and for $p_D^\mathcal{A}=0.5$. Here, the experimental data is given by the blue triangles.  All error bars indicate one sigma error, as determined based on experimental measurements (see Appendix E).}
\end{center}
\end{figure}

\paragraph*{Thermodynamic considerations.} The engine cycle we have designed and implemented experimentally is seen to pump energy to the quantum load in a cyclic manner. While this is useful in its own right, one might wonder whether the load is merely being heated up. To answer this question, we go one step further and study whether the load can function as an effective quantum battery \cite{CampaioliPollock2018, AndolinaPolini2018b}. 
In other words, we ask whether and how much energy can be extracted from the load with unitary operations. 
To this end it is useful to introduce the concept of {\it passive} states \cite{PuszWoronowicz1978, allahverdyan2004maximal}. The defining quality of a passive state, such as a thermal state, is that no work can be extracted from it. A state is passive when the associated density matrix is diagonal in the representation of energy eigenstates, with the diagonal terms (the occupation probabilities) decreasing for increasing energy eigenvalues. The state in Fig.\ref{fig:fig3}(a) is, for example, passive. Applying this criterion to analyze our quantum load, the energy which is extractable using unitary operations is given by the {\it ergotropy} $\mathcal{E}_L$, which is defined to be the difference in energy between the state of the load, $\rho_L(t)$, and the so-called ``passified'' state, $\tilde{\rho}_L(t)= U \rho_L(t) U^{\dagger}$ where $U$ is a unitary transformation which makes the state $\rho_L(t)$ passive \cite{allahverdyan2004maximal}. More precisely, $\mathcal{E}_L(t)=\tr[H_L \rho_L(t) ]-\tr[H_L\tilde{\rho}_L(t)]$. To investigate the ergotropy of the load we look at the distribution of $p_n$ within the load, and make the assumption (justified below), that the off-diagonal elements of the density matrix of the load $\rho_L(t)$ may, in our case, be neglected when evaluating the ergotropy. Note that at the time at which the ergotropy is experimentally measured, the engine and the load are decoupled.

Fig.~\ref{fig:fig3}(a,b) shows two examples of $p_n$ distributions, extracted based on fitting to experimental data.  The initial state of the load [Fig.~\ref{fig:fig3}(a)] is thermal, and hence passive.  After $N_c=8$ cycles [Fig.~\ref{fig:fig3}(b)], the occupation probability profile  is no longer passive and hence the load has nonzero ergotropy. This can be seen from the shift in the location of the peak from $n = 0$ in [Fig.~\ref{fig:fig3}(a)], to $n = 4$ in [Fig.~\ref{fig:fig3}(b)]. 
Theoretically, under ideal conditions of perfect state transfer and in the absence of dissipation, the ergotropy would grow with increasing number of cycles as $\mathcal{E}_L \propto$  $N_c-\alpha\sqrt{N_c}$, where $\alpha$ is a constant (see Appendix F). 
The detailed time dependence of the ergotropy measured in the experiment is shown in Fig.~\ref{fig:fig3}(c) by the green $\times$. The ergotropy clearly grows with the number of cycles, signaling that successive iterations of the same cycle are able to continuously increase the ergotropy of the load. This ergotropy is evaluated assuming a diagonal reduced density matrix of the load, we thus numerically confirm that this is a valid approximation by showing the proximity of the numerical values for the approximated ergotropy (dashed black line), and the exact ones (continuous red line). The numerical and experimental values are in very good agreement.

In addition to the change in ergotropy, Fig.~\ref{fig:fig3}(b) shows that the variance in the phonon number increases over a few cycles. This behavior is consistent with a pattern of biased diffusion, as predicted in an earlier theoretical study \cite{teo2017converting}. 
Figs.~\ref{fig:fig3}(d-e) show both numerical and experimental results for the evolution of the von Neumann (information) entropy $\mathcal{S}_L = -\tr\left[\rho_L \log(\rho_L)\right]$, for an engine-load coupling implemented by the resonant sideband transfer protocol. The red continuous line represents the exact information entropy of the load while the black dashed line represents an estimate of the information entropy neglecting the off-diagonal elements of the reduced density matrix of the load. Since the two curves are very close to each other, the experimentally measured values of $p_n$ can be used to estimate the entropy of the load, depicted by blue squares and triangles in Figs.~\ref{fig:fig3}(d,e). Fig \ref{fig:fig3}(d) shows the entropy increasing over $N_c=8$ cycles, for $p_D^\mathcal{A}=0.32$. Fig.~\ref{fig:fig3}(e) shows the entropy descreasing over the course of $15$ cooling cycles, with $p_D^\mathcal{A}=0.5$. This change of the entropy of the load over the course of the cycles is a clear signature of the presence of correlations between the engine and the load.

\paragraph*{Conclusions.} This work describes the experimental implementation and theoretical study of both engine and refrigerator quantum cycles, using two electronic states of a single trapped ion as the working fluid, and one vibrational mode of the same ion as the quantum load. We demonstrate that when engine cycles are run, the ergotropy, entropy, and average phonon number in the load increase. When the device is operated as a refrigerator, these three quantities are shown to decrease. We also run the device under different forward-cycle conditions, demonstrating the capability of variable tuning of its functioning. To our knowledge, this is the first cyclic quantum energy conversion device operating with {\it strong} coupling to a load which operates in the deep quantum regime, and in the presence of non-trivial back action from the load. 
Other peculiar features of our energy-conversion device include (i) the resetting of working fluid to nonequilibrium states, due to the use of both coherent laser excitation and dissipative optical pumping in implementing the strokes, and (ii) the gigantic energy scale mismatch between the working fluid and the quantum load. 

A detailed analysis of the dynamical and thermodynamic properties of the engine cycles shows that, in both theory and experiment, the engine-load coupling results in significant correlations between the engine and the load, and non-trivial information entropy generation and flow between them. It is thus remarkable that the load can still operate as an effective quantum battery. 
Good agreement between theoretical modeling and experimental results confirms our understandings presented above.

We briefly comment on the efficiency of the engine cycle, which couples optical states of the working fluid with the vibrational states of the load.  Energy-wise, it may not be suitable to treat these two degrees of freedom on the same footing. It is thus relevant to define an efficiency of conversion of quanta, $\eta_Q$, as the net increase in the mean phonon number in the load, divided by the total input of photon quanta in the working fluid throughout the various strokes.
For the cycle with $p_D^\A=0.32$ and a resonant sideband transfer for strokes (II) and (IV), as in Fig.\ref{fig:fig1}(c), the occupation in $p_D^\B\approx 0.52$ and in $p_D^\D\approx 0.58$, while the average increase of the mean phonon number per cycle is $\approx 0.4$. This results in $\eta_Q \approx 0.4/1.1\approx 0.36$. For larger $p_D^\A$ this efficiency increases, and in particular, for $p_D^\A=1$ it can reach $\eta_Q \lesssim 1$.

Given the importance of measurements in quantum mechanics, and specifically in quantum thermodynamics, we conclude with a comment on the role of measurements. A measurement made on the state of the engine collapses both the states of the engine and the load.  This makes it important to specify how measurements are performed in our experiment. In this work, the device is run for various numbers of cycles and then a single measurement is performed at the end of the prescribed number of cycles. 
However, we stress that, in our set-up, an unselective measurement of the energy of the load \cite{WatanabeHanggi2014, watanabe2017quantum} at an intermediate or later times would have little impact on the evolution of the system for two reasons: First, the engine and the load are in a product state at the end of strokes (I) and (III).  
Second, even if quantum coherence can build up in both the engine and the load during the cycle, as explained above, because of the inherent decoherence in the experiment, the quantities we study are well described by a diagonal density matrix. 
Further experiments are needed to gain deeper insights into the role of measurement in quantum energy-conversion devices.

\paragraph*{Acknowledgments} Numerical simulations were carried out using QuTiP \cite{QuTiP, QuTiP2}. J.G., M.M., D.P. and P.H acknowledge support from the Singapore MOE Academic Research Fund Tier-2 project (Project No. MOE2014-T2-2-119 with WBS No. R-144-000-350-112). D.Y. acknowledges the support by the National Research Foundation Singapore under its Competitive Research Programme (CRP Award No. NRF-CRP14-2014-02) and T.D. acknowledges the support from the Singapore MOE Academic Research Fund Tier-2 project (no. MOE2016-T2-2-120). N.V.H, M.M., and D.P. acknowledge the work of A. Honda in creating the cycle image.

\bibliography{References3}
\bibliographystyle{apsrev4-1}

\begin{appendix}
\setcounter{figure}{0}  
\renewcommand{\thefigure}{A\arabic{figure}}

\section{Experimental setup}
Our energy-conversion device consists of a singly-ionized $^{138}$Ba$^+$ atom in a linear Paul trap, with a radial trap frequency of $\sim 1.7~$MHz. The $^{138}$Ba$^+$ ion has $5$ internal energy levels relevant to this experiment. Four lasers at wavelengths of $493$~nm, $650$~nm, $614$~nm and $1,762$~nm are needed to address these states. The $493$~nm and $650$~nm lasers are used for Doppler cooling and re-pumping out of the D$_{\frac{3}{2}}$ level, as well as for quantum state detection by fluorescence observation. During these processes, the Zeeman splitting is not resolved. The $614$~nm laser is used in conjunction with the $1,762$~nm laser for further sideband cooling, nearly to the ground state of the radial external motional mode. A separate, circularly polarized $493$~nm laser, along with the $650$~nm laser, is used to initialize the ion to the ground state of the engine, S$_{\frac{1}{2},-\frac{1}{2}}$, by optical pumping~\cite{dehmelt1957slow}. The $1,762$~nm laser is also used to address the two energy levels which constitute the \textit{engine} of our device, namely the S$_{\frac{1}{2},-\frac{1}{2}}$ and the D$_{\frac{5}{2},-\frac{5}{2}}$ states. This transition can alternatively be described as our quantum bit (i.e. ``qubit'')~ transition, due to its long ($\sim 30$ second) lifetime. This is consistent with the negligible decay of the D$_\frac{5}{2}$ state in our experimental setup when observed over more than $2$~s in absence of any external fields. Technical details of the lasers themselves and laser locking setups are available in \cite{DeMunshiMukherjee2015, DeMunshiMukherjee2016}. The $650$~nm and $614$~nm lasers, along with the Doppler-cooling $493$~nm laser, are combined into a single beam before being focused on the ion, while the $1,762$~nm and $493$~nm (optical pumping) beams are applied separately, through different optical viewports into the ultra-high vacuum chamber. The $1,762$~nm laser's frequency is tuned by generating a signal either using a direct digital synthesizer (DDS), or an arbitrary waveform generator (AWG), for resonant sideband or adiabatic sideband transitions, respectively.  The signal produced by the DDS or AWG is channeled by a switch, and then amplified before going to an acousto-optic modulator (AOM).

To implement the simplified two-level system depicted in Fig.\ref{fig:fig1}, three pairs of coils in the Helmholtz configuration are used to apply a $4~$Gauss external magnetic field along the axis of the $493$~nm (optical pumping) laser beam. This Zeeman-splits the otherwise degenerate internal atomic energy levels, and defines an axis of quantization of the atom. To initialize the ion to the well-defined ground state, S$_{\frac{1}{2},-\frac{1}{2}}$,~ at the beginning of each experiment we apply the $493$~nm optical pumping beam, with left circular polarization, propagating along the quantization axis. 

As mentioned previously, the engine of the device is formed by the approximate two-level system consisting of the S$_{\frac{1}{2},-\frac{1}{2}}$ and the D$_{\frac{5}{2},-\frac{5}{2}}$ states. The transition between these states is addressed by the narrow line-width (nearly single frequency) $1,762$~nm laser (linewidth $\lesssim 100~$Hz). The two-level approximation holds since the $1,762$~nm laser hardly shifts any other energy level of the atom while addressing the S$_{\frac{1}{2},-\frac{1}{2}}$ and D$_{\frac{5}{2},-\frac{5}{2}}$ levels on resonance.

The ion's internal quantum state, i.e. (for our purposes) whether it is in the S$_{\frac{1}{2},-\frac{1}{2}}$ or D$_{\frac{5}{2},-\frac{5}{2}}$ state, is determined by measuring its fluorescence while exciting it with only the $493$~nm and $650$~nm lasers. If the ion is in the D$_{\frac{5}{2},-\frac{5}{2}}$ state, it is not affected by the $493$~nm and $650$~nm lasers, and no fluorescence is observed. If the ion is not in the in the D$_{\frac{5}{2},-\frac{5}{2}}$, exposing it to the $493$~nm and $650$~nm lasers causes it to excite and de-excite continuously on the $493$~nm transition, emitting $493$~nm photons. In this way, by measuring $493$~nm photons one can distinguish between the ground state and the excited state with nearly $100\%$ efficiency. Repeating such a measurement many times, and averaging the results, yields a value $p_S$, i.e. the probability that the ion is in the $S_{\frac{1}{2},-\frac{1}{2}}$ state.

\section{Experimental procedure}

The ion is first Doppler cooled for $300~\mu$s using the $493$~nm and the $650$~nm lasers, and then sideband cooled for $\sim 25$~ms by continuously applying the red-detuned $1,762$~nm and the $614$~nm lasers.  One full cycle of the device is implemented via the following steps:

\begin{enumerate}[label=({\Roman*})]

\item The first step, $\D$ to $\A$ in Fig.~\ref{fig:fig1}(c), consists of two parts:
\begin{enumerate}[label=(\subscript{I}{{\alph*}})]
\item (Setting the engine to its ground state) Any population in the D$_{\frac{5}{2},-\frac{5}{2}}$ state is transferred from the D$_{\frac{5}{2},-\frac{5}{2}}$ state to the S$_{\frac{1}{2},-\frac{1}{2}}$ state via the P$_{\frac{3}{2}}$ state (without resolving the Zeeman splitting).  This is done by simultaneous application of the $614$~nm, $650$~nm, and $493$~nm optical pumping lasers for $\sim 5$ to $50\mu$s. The $493$~nm optical pumping beam uses circularly-polarized light to continuously empty the $S_{\frac{1}{2},+\frac{1}{2}}$ state. 

\item (Resetting the engine to state $\A$) The $1,762$~nm laser is applied to the ion at exactly the carrier transition frequency, for somewhere between $0$ and $2.1 ~\mu $s, depending on the desired value of ~$p_D^\A$. (Note: $2.1 ~\mu $s corresponds to half a period of the carrier Rabi oscillation, see Appendix D). 
\end{enumerate}

\item (Red sideband transfer, $\A$ to $\B$) A red-sideband transfer is performed by applying the red-detuned $1,762$~nm laser for $180~\mu$s (resonant transfer), or by sweeping ($\pm ~30~$kHz) across the red-sideband frequency, over $4$~ms, in the case of the rapid adiabatic protocol. Here, red-sideband and red-detuned refer to the wavelength given by $\lambda_{r.s.} =$ ($1,762$~nm carrier transition) $ + 2\pi c/\omega$, where $c$ is the speed of light and $\omega$ is the radial motional frequency of the ion in the trap, $2\pi \times 1.7$~MHz.

\item (Setting the engine to its ground state, $\B$ to $\C$) Step (III) is the same as step (I${_a}$).

\item (Blue sideband transfer, $\C$ to $\D$) A blue-sideband transfer is performed by applying the blue-detuned $1,762$~nm laser (i.e. $\lambda_{b.s.} = $ ($1,762$~nm carrier transition) $ - 2\pi c/\omega$),~ for $180~\mu$s for the resonant transfer, or by sweeping ($\pm ~30~$kHz across the blue-sideband frequency, over $4$~ms, for the rapid adiabatic protocol.

\end{enumerate}

These strokes are repeated an integer number of iterations ranging from $N_c=0$ to $8$ when using resonant sideband transfer for the coupling between the two-level system and the harmonic oscillator, or $N_c=0$ to $5$ when using the rapid adiabatic transfer protocol. In stroke (I) the engine is prepared in state $\A$, in stroke (II) the engine and the load are coupled, then, in stroke (III), the engine is prepared in state $\C$, and in stroke (IV) the engine and load are coupled again.

After the intended number of cycles is run, both the internal and external states of the ion are detected by performing a blue-sideband Rabi excitation scan. Here $\lambda_{b.s.}$ is applied for times ranging from $t = 0$ to $t \ge 100~\mu$s, and at each time value the state of the ion is detected. For each data point, the full set of steps described above is repeated $150$ to $200$ times, and the fraction of the outcomes where the ion is found in the $S-$state is calculated. 

Thus, to obtain the final state of the engine and the load after a given number of cycles, a measurement such as shown in Fig.\ref{fig:fig5} requires $30,000$ experiments.

\section{A graphical representation of the cycle}

\begin{figure}
\begin{center}
\includegraphics[width=1\columnwidth]{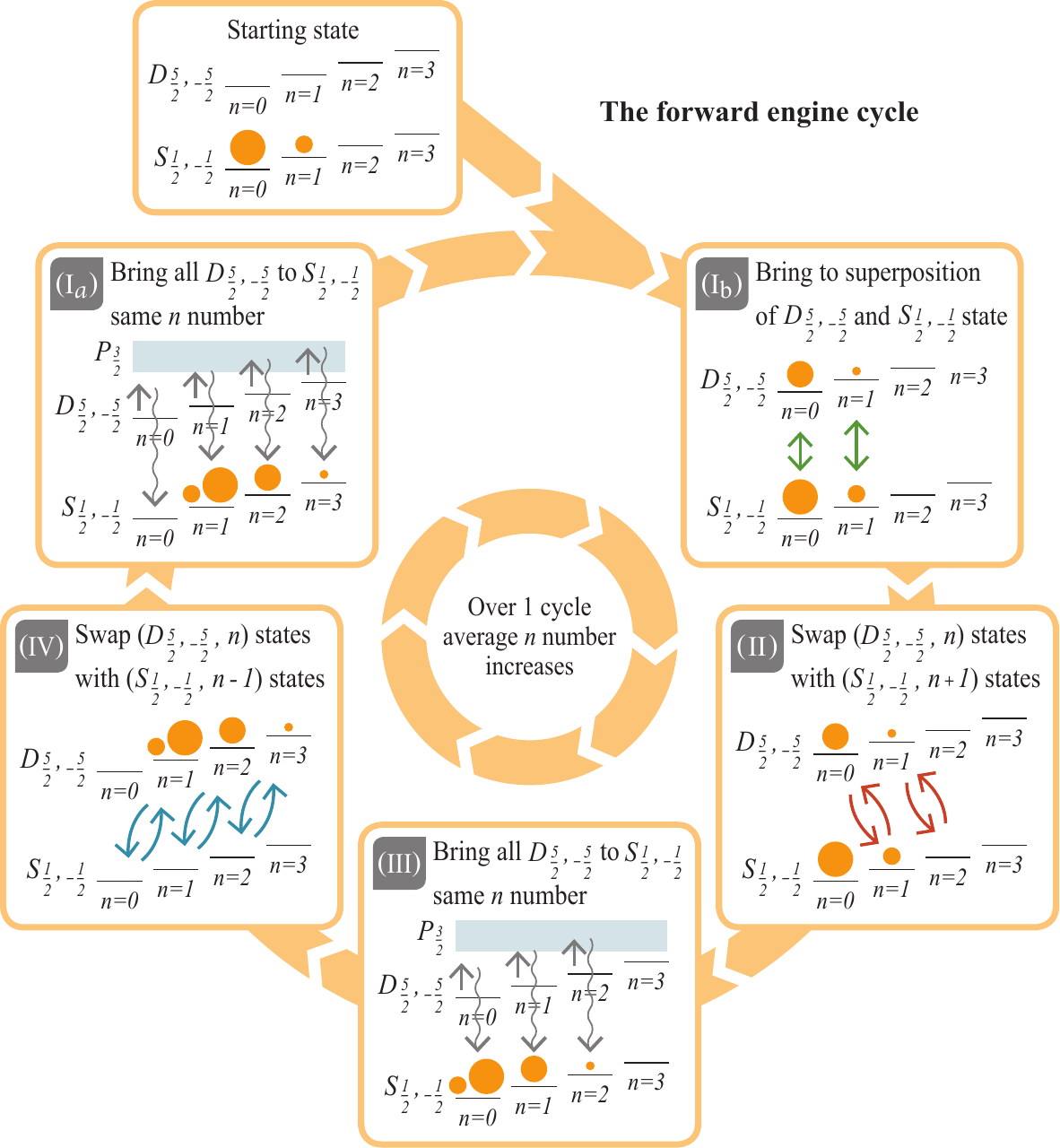}
\caption{\label{fig:fig4} Pictorial description of the forward engine cycle. The initialized state is shown, followed by strokes (I$_b$) through (IV), and finally back to (I$_a$).  The diameter of the orange circles represents the probability of the ion being in the corresponding state.}
\end{center}
\end{figure}

Fig.\ref{fig:fig4} provides an intuitive understanding of the forwards engine cycle. The system we study is composed of a two-level system and a harmonic oscillator, which can be represented in the basis $|s,n\rangle$.  The notation $s = \uparrow, \downarrow$ is used to indicate, respectively, the excited and the ground state of the two-level system, while $n$ represents the level of the harmonic oscillator. The energy difference between the two levels $s =\; \uparrow, \downarrow$  is much larger than between two levels of the harmonic oscillator. For clarity, the allowed energy states of the system can be spread out horizontally and represented as two ``staircases'' on top of each other. The bottom staircase corresponds to the lower energy state of the two-level system, (S$_{\frac{1}{2},-\frac{1}{2}}$), while the top staircase corresponds to the upper energy state, (D$_{\frac{5}{2},-\frac{5}{2}}$). In each staircase, different steps correspond to different levels $n$ of the harmonic oscillator, increasing from left to right. The diameter of the orange circles schematically represents the probability of the ion being in the corresponding energy state.

The engine cycle is composed of $4$ strokes, shown clockwise from (I) to (IV) in Fig.\ref{fig:fig4}. The top left box (without a number), shows the initial state. In the initial state, the two-level system is in the ground state, and the harmonic oscillator has the largest probability of the ion being in the ground state ($n=0$ of the harmonic oscillator), and a decreasing probability for the higher levels. The cycle then proceeds to the second part of the first stroke, (I$_b$). Applying the $1,762$~nm laser for a predetermined amount of time brings the engine to a superposition of its two levels. In stroke (II), a Jaynes-Cummings coupling moves the population from state $| \downarrow, n+1\rangle$ to $| \uparrow, n\rangle$, and the population from state $| \uparrow, n\rangle$ to $| \downarrow, n+1\rangle$, as represented by the red arrows. For the values of ~$p_D^\mathcal{A}$~ studied, this stroke results in a decrease in the average $n$ number, $\langle n \rangle_L$, of the harmonic oscillator. In stroke (III) the two-level system is no longer coupled to the harmonic oscillator, and it is reset to its ground state by a dissipative process. Stroke (IV) is similar to stroke (II), but here an anti Jaynes-Cummings interaction is used between the two-level system and the harmonic oscillator.  For this case, ($| \downarrow, n\rangle$ is swapped with $| \uparrow, n+1\rangle$, and viceversa (as represented by the blue arrows). This step produces the most significant increase in the average phonon number, $\langle n \rangle_L$, of the harmonic oscillator. Finally, stroke (I$_a$) resets the two-level system to its ground state, with the two-level system decoupled from the harmonic oscillator. We note that the physics leading to the redistribution of the harmonic oscillator population can be captured in steps (I) and (II).

\section{Modeling of the experiment}   

Each stroke $M$ is modeled by a weak-coupling Markovian master equation of the form  
\begin{align}
\frac{d\rho_{_{E+L}}}{dt}  = & -\frac{\im}{\hbar}[H_E+H_L+V_M(t),\;\rho_{_{E+L}}] \nonumber \\ 
& + \mathcal{D}^{L}_{\phi,M}(\rho_{_{E+L}}) + \mathcal{D}^{L}_{h,M}(\rho_{_{E+L}}) + \mathcal{D}^{E}_{M}(\rho_{_{E+L}}).     
\end{align}
Here, $\rho_{_{E+L}}$ is the density matrix of the engine$+$load system, $H_E$ is the  Hamiltonian of the engine, $H_L$ is the Hamiltonian of the load, $V_M(t)$ is the interaction Hamiltonian which characterizes the strength of the interaction between the system and the load, for the strokes $M=($II$)$ and $($IV$)$ and $[\cdot , \cdot]$ denotes the commutator. The three main dissipative processes are described by $\mathcal{D}^{L}_{\phi,M}(\rho_{_{E+L}})$, for the dephasing of the vibrational mode (the load) due to fluctuations in the trap RF power, $\mathcal{D}^{L}_{h,M}(\rho_{_{E+L}})$ for heating of the vibrational mode due to patch potentials, and $\mathcal{D}^{E}_{M}(\rho_{_{E+L}})$ for the decay of the two-level atom (the engine) due to spontaneous emission \cite{PetruccioneBreuer2002,LeibfriedWineland2003}. In more detail, $\mathcal{D}^{L}_{\phi,M}(\rho_{_{E+L}})$ is given by   
\begin{align}
\mathcal{D}^{L}_{\phi,M}(\rho_{_{E+L}}) = \gamma^{L}_{\phi,M} \left(n \rho_{_{E+L}} n -1/2\left\{n,\rho_{_{E+L}}\right\} \right) \nonumber \\ 
\end{align}
where $\gamma^{L}_{\phi,M}$ is the dephasing rate of the vibrational mode, $n$ is the phonon-number, and the brackets denote the anti-commutator. $\mathcal{D}^{L}_{h,M}(\rho_{_{E+L}})$ is given by     
\begin{align}\label{eq:heating}
\mathcal{D}^{L}_{h,M}(\rho_{_{E+L}}) & = \gamma^{L}_{h,M} n_{m}\left(a^\dagger \rho_{_{E+L}} a -1/2 \left\{a a^\dagger,\rho_{_{E+L}}\right\} \right) \nonumber \\ 
& + \gamma^{L}_{h,M} (1+n_{m})\left(a \rho_{_{E+L}} a^\dagger -1/2 \left\{a^\dagger a,\rho_{_{E+L}}\right\} \right). \nonumber \\ 
\end{align}
Here, $\gamma^{L}_{h,M}$ is the heating rate of the vibrational mode, $n_m$ is the target occupation of the mode, towards which Eq.(\ref{eq:heating}) drives the harmonic oscillator, while $a^\dagger$ and $a$ are the creation and annihilation operators for the harmonic oscillator, respectively. $\mathcal{D}^E_{M}(\rho_{_{E+L}})$ is given by    
\begin{align}
\mathcal{D}^E_{M}(\rho_{_{E+L}}) & = \gamma^{E}_{M} \left(\sigma^- \rho_{_{E+L}} \sigma^+ -1/2 \left\{\sigma^+\sigma^-,\rho_{_{E+L}}\right\} \right)       
\end{align}
where $\gamma^{E}_{M}$ is the spontaneous emission rate of the excited electronic state of the atom and $\sigma^\pm$ are the raising and lowering operators for the electronic states \cite{PetruccioneBreuer2002}.

Throughout a given stroke $M$, we have the parameter of the engine Hamiltonian $\nu=170~$THz (the frequency difference between the two levels), while for the load Hamiltonian $\omega=1.7~$MHz (the oscillation frequency of the ion). The coupling between the engine and the load is set to zero both for the first and the third strokes, i.e. $M=($I$),\;($III$)$.

The dissipation rates for the load have been measured in separate experiments to be $\gamma^L_{\phi,M} = 318~$Hz and $\gamma^L_{h,M} \le 1$Hz (we use $\gamma^L_{h,M} =0.4~$s$^{-1}$ in our simulations). The atomic decay rate for the engine is measured to be $\ll 1$Hz, so we use $\gamma^L_{h,M} =0.4~$s$^{-1}$ in the numerics.  
The dominating decay is thus the dephasing of the motional oscillation while the motional heating rate of the ion in the trap is consistent with the size of the trap. These parameters are fixed in all strokes except the optical pumping stroke. 
In strokes (I$_a$) and (III), the optical pumping to the ground state has been modeled by using $\gamma^E_{{\rm I}}=\gamma^E_{{\rm III}}\approx700~$kHz. Experimentally this is achieved by transferring the population to a third excited state which spontaneously decays at a similar rate as modeled.
In stroke (I$_b$), an extra term $\hbar\Omega_0\sigma^x$, where $\sigma^x=\sigma^++\sigma^-$, is added to the engine Hamiltonian to operate a Rabi transfer between the $S$ and $D$ levels. The on-resonance coupling strength used in our experiment is measured to be $\Omega_0 = 121.7$kHz. This results in an evolution of the excited $D$ state as a function of time given by $p_D = e^{-\gamma_{E,M}\Delta t} \sin^2(\Omega_0 \Delta t/2)$ \cite{LeibfriedWineland2003}. 
In order to set the desired population ratio for the two-level system, $\Delta t$ is adjusted while $\gamma^{E}_{M}$ and $\Omega_0$ are fixed by their experimental values.      

Strokes (II) and (IV) are executed either via a resonant sideband Rabi pulse, or a sideband rapid adiabatic passage protocol. The only difference between strokes (II) and (IV) is the frequency of the applied laser, which either corresponds to the first red-sideband, which couples $|\dw,n\rangle$ with $|\uw,n-1\rangle$, or to the first blue-sideband, which couples $|\dw,n-1\rangle$ with $|\uw,n\rangle$. For the resonant sideband transfer, the coupling between the engine and the load is given by 
\begin{align}
V(t)^{res}_{({\rm II})} = \hbar g_{n,n'}(t)(\sigma^+a + \sigma^-a^\dagger) \nonumber \\ 
V(t)^{res}_{({\rm IV})} = \hbar g_{n,n'}(t)(\sigma^+a^{\dagger} + \sigma^-a)
\end{align}
where $g_{n,n'}$ is the sideband Rabi frequency when the states $|\dw,n\rangle$ and $|\uw,n'\rangle$ are coupled. The sideband Rabi frequency $g_{n,n'}$ is given by $g_{n,n'} = \eta\sqrt{n_>}\Omega_0(t)$, where $n_>$ denotes the greater of the two values, ${n}$ and ${n'}$, $\eta=0.012$ is the Lamb-Dicke parameter and $\Omega_0$ is the sideband Rabi frequency for the transition from $|\dw,0\rangle$ and $|\uw,1\rangle$. The time dependence in $\Omega_0(t)$ comes from the applied square pulse of the laser inducing the Rabi frequency at a constant intensity. Thus, letting $t_M$ be the duration of the stroke $M=($II$),\;($IV$)$, we use $t_M = \pi/g_{0,1}$ for a complete population inversion when $n=0$. 
Since the coupling strength $g_{n,n'}$ depends on $n$, as $t_M$ is fixed, the percent of the population transferred, for a given time $t$, is lower for higher $n$ values.

We also performed experiments using the rapid adiabatic passage (RAP) protocol, which has a weaker theoretical dependence on $n$. The coupling in the RAP interaction Hamiltonian is of Landau-Zener type, and it acts simultaneously on multiple levels $n$ of the load. It is given by 
\begin{align}
V(t)^{RAP}_{({\rm II})} = \hbar \Delta(t)\sigma_z + 2\pi \hbar g_0 (\sigma^+a + \sigma^-a^\dagger) ,\\
V(t)^{RAP}_{({\rm IV})} = \hbar \Delta(t)\sigma_z + 2\pi \hbar g_0 (\sigma^-a + \sigma^+a^\dagger),
\end{align}
where $g_0=\eta\Omega_0\approx1.5$kHz. Here, $\Delta(t) = -\Delta_0(1-2t/\tau)$. $\Delta_0$ is the range of frequencies across which the laser is swept, on either side of the sideband transition. $\sigma_z$ is the Pauli matrix, and $\tau$ is a parameter which is used to determine the sweep rate. The sweep rate is dictated by the ratio $\Delta_0/\tau$.  For the simulations, we use $\Delta_0 = \pm 30~$kHz~ and $\tau=4.0$ms. These values are based on results obtained from many experiments, performed while varying $\Delta_0$ and $\tau$ to optimize the RAP transfer protocol for maximum transfer efficiency. In both strokes (II) and (IV), we achieve maximum transfer efficiency rates of $\sim80\%$. 
 
To simulate steps (I) through (IV) of the engine cycle, the time-dependent Hamiltonian is solved numerically using the python QuTiP package. This gives the density matrix of the engine and the load at each point in time. The density matrix, $\rho_{E+L}$, is then used to evaluate the observables of interest, to extract results from the observables and to compare to experimental measurements.

\section{Data analysis}
\begin{figure}[h!]
\begin{center}
\includegraphics[width=1\columnwidth]{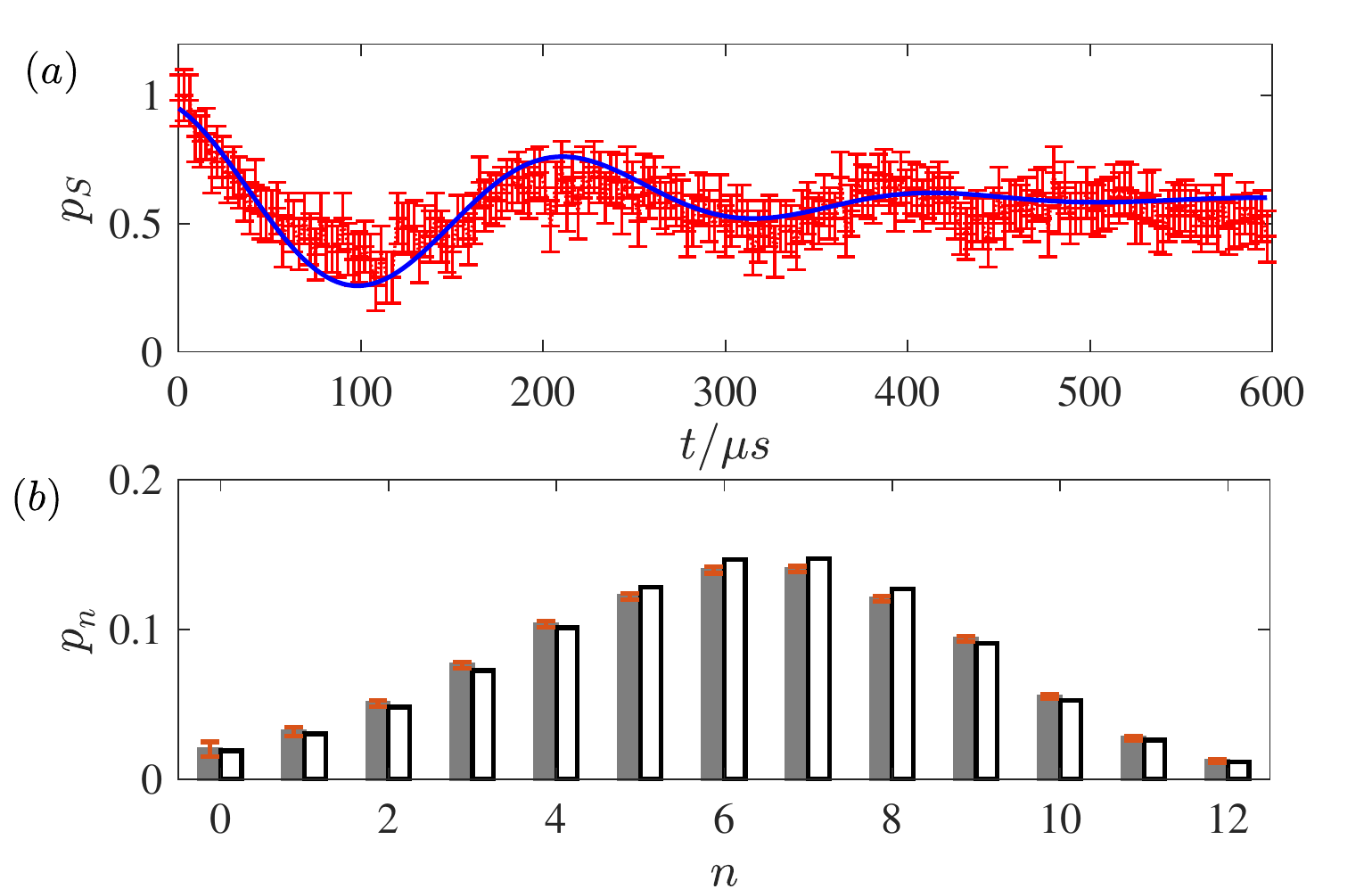}
\caption{{\label{fig:fig5}(a) The population of the $S-$state $p_S$ as a function of the exposure-time of the ion to the blue sideband laser excitation. The blue line is a fit using Eq.~(\ref{eq:mod}) with reduced $\chi^2=0.7$. (b) Distribution of occupation, $p_n$, of the different levels of the harmonic oscillator mode after $N_c=8$ cycles. The distribution obtained from numerical calculation (empty bars) is compared to the one obtained by the fit to the experimental data (shaded bars). }%
}
\end{center}
\end{figure}
The heat engine, or refrigeration operations are implemented for a certain number of complete cycles ($N_c=2$, $4$, $6$, $8$ or $15$ for resonant transfer and $N_c=1$, $3$ and $5$ for the RAP transfer). At the end of a given number of cycles, the blue-sideband laser is applied for a duration $t$. Then, a measurement on the ($S$-state to $P$-state) transition gives an outcome indicating whether the ion is in the $S$-state, or the $D$-state. This procedure is repeated $150$ times for each duration of the application of the blue-sideband laser. The $150$ experiments are then averaged to yield the \textit{probability} that, given the operations to which the ion was previously subjected (the cycles), it is in the $S-$state upon measurement. This probability is called $p_S$. $p_S$ is then measured for different excitation times, ($t + 3\mu$s, $t+6\mu$s... etc.), with $150$ experiments each time.  For a typical scan, this process is repeated up to $t \geqslant 100~\mu$s. For specific cases, to establish reference values, scans are run up to $t = 600~\mu$s. An example of a reference curve, along with the experimental 1-sigma error, is shown in Fig.\ref{fig:fig5}(a).  This curve corresponds to the case $N_c=8$ and $p^\A_S = 0.5$.  Outlier data points, $p_S$, were removed manually by visually monitoring scans as they were in progress and noting the time when the photo-multiplier-tube (PMT) detector signal dropped suddenly (due to occasional collisions), or increased suddenly (due to occasional laser instabilities).  

For each number of cycles $N_c$, the final state of the load is estimated from a scan such as just described, using the protocol outlined in \cite{PhysRevLett.76.1796}.  Theoretically, the $p_S(t)$ due to resonant sideband excitation is a linear function of the initial population distribution in the load (harmonic oscillator).  The probability of the ion being in a given level of the harmonic oscillator, $p_n$, is related to $p_S$ by:
\begin{equation}
p_S = \sum_n p_n\cos^2(\Omega_{n,n+1} t)e^{-\gamma_n t},\label{eq:mod}
\end{equation}
where $\Omega_{n,n+1}$ and $\gamma_n$ are the blue sideband Rabi frequency and the decay rate for the $n-$th motional state \cite{LeibfriedWineland2003}. $\Omega_{n,n+1}$ is determined from the experimental value of the $\Omega_{0,1}$.  For $\gamma_n$ we use the experimentally measured value $\gamma_n = \gamma_{h,M}^L\sqrt{n+1}$. In order to extract the distribution that is most likely to describe the experimental result, a least-squared fit is performed on the data using the model curve described by Eq.~(\ref{eq:mod}). The fit is constrained using the minimum number of $n$ values which accurately capture the distribution. Typically, this is around $12$ to $15$ $n$ values. As a starting distribution, the fit routine uses the expected distribution obtained from numerical simulations. The fit is then restricted to search for values of $p_n$ which are within $\pm5\%$ of the numerical simulation results. In most cases, the resulting reduced $\chi^2$ value is below $1$, suggesting an overestimate of the phase error in the experiment. Our estimate of the phase error is based on over 150 repetitive measurements of $p_S$. In Fig.~\ref{fig:fig5}(a), the blue continuous line shows the fit for the corresponding data.

Fig.~\ref{fig:fig5}(b) shows the distribution of $p_n$ derived from numerical simulation (empty bars), as well as the fitted distribution of $p_n$ (shaded bars). The error on the fitted values of $p_n$ is calculated using a standard technique of error estimation for parameters of a least-squares fit \cite{Bevington1969}. We note that to reduce the running time of the overall experiment, we have only produced reference scans for $N_c = 0$ and $N_c = 8$ cycles. The number of data points in these scans was chosen to be up to $8$ times greater than for other numbers of cycles. Therefore, for these scans the uncertainties on the derived observable are correspondingly smaller (Fig. 2(a-c)).

\section{Ideal evolution of the engine and load}
To develop a clearer understanding of how the energy-conversion device works, it is useful to consider its evolution under ideal conditions. In this section, we consider (i) that the lasers do not induce any dephasing, (ii) that there are no spontaneous emissions, and (iii) that transfers in strokes (II) and (IV) have an efficiency of $100\%$ which is independent of the $n$ level. We begin our analysis with both the engine and the load in their ground state and, to simplify the explanation, we start our analysis of the cycle with stroke (IV). The initial condition is $\rho_{_{E+L}}^{\mathcal{C}_0} = | \dw, 0 \rangle \langle \dw,0 |$~, where $\rho^{\mathcal{C}_0}_{_{E+L}}$ is the density matrix before the blue sideband operation (which takes the ion from $\mathcal{C}$ to $\mathcal{D}$ in Fig. 1(c)). The subscript ``$0$'' refers to the number of times that the operation ($\mathcal{C}$ to $\mathcal{D}$) has been implemented \textit{prior} to the given density-matrix expression. 

We now proceed by implementing stroke (IV) of the cycle. The effect of the AJC coupling is to shift the engine to the $D_{\frac{5}{2},-\frac{5}{2}}$ state, while simultaneously transferring $n$ to $n+1$. Therefore, stroke (IV) gives
\begin{align}
\rho^{\mathcal{D}_1}_{_{E+L}} = | \uw, 1 \rangle \langle \uw,1 |.
\end{align}
We now apply stroke (I). For the sake of giving a generic example, we consider the case in which after step (I${_b}$), ~$p_D^{\mathcal{A}}=1/4$. Stroke (I) produces a pure state
\begin{align}
\rho^{\mathcal{A}_1}_{_{E+L}} = \frac{1}{4}\left(|\uw\rangle+\sqrt{3}e^{\im\phi}|\dw\rangle\right)\left(\langle \uw | + \sqrt{3}e^{-\im\phi}\langle \dw |\right) \otimes \proj{1},
\end{align}
where $\rho^{\mathcal{A}_1}_{_{E+L}}$ is the density matrix at point $\mathcal{A}_1$. Here, $\phi$ is a phase which can be determined experimentally, and that can be fixed such that it returns to the same value at the end of any integer number of cycles. Now we proceed to stroke (II) of the cycle, an ideal JC coupling which transfers $|\dw,n\rangle$ to $|\uw,n-1\rangle$ and viceversa. Stroke (II) thus results in an entangled state
\begin{align}
\rho^{\mathcal{B}_1}_{_{E+L}} = \frac{1}{4}\left(|\dw,2\rangle+\sqrt{3}e^{\im\phi}|\uw,0\rangle\right)\left(\langle \dw,2 | + \sqrt{3}e^{-\im\phi}\langle \uw,0 |\right).
\end{align}
Finally, stroke (III) brings the engine and the load to the state
\begin{align}
\rho^{\mathcal{C}_1}_{_{E+L}} = \frac{1}{4}\proj{\dw}\otimes(3\proj{0} + \proj{2}).
\end{align}
Note that after the four strokes, the engine is back to its initial state, $|\dw\rangle\langle \dw|$, but the load has a higher average occupation, $\langle n \rangle_L$. Furthermore, the load goes from a pure state, to a mixed state. Continuing from $\rho^{\mathcal{C}_1}_{_{E+L}}$ and progressing through an additional cycle, the reduced density matrix of the load evolves as 
\begin{align}
\rho^{L,\mathcal{D}_2}_{_{E+L}} &= \rho^{L,\mathcal{A}_2}_{_{E+L}} = \frac{1}{4}(3\proj{1} + \proj{3})  \\ 
\rho^{L,\mathcal{B}_2}_{_{E+L}} &= \rho^{L,\mathcal{C}_2}_{_{E+L}} = \frac{1}{16}(9\proj{0} + 6\proj{2} + \proj{4}).
\end{align}
For the given parameter value, $p_D^{\mathcal{A}}=1/4$, it can be shown that the change in the average phonon number of the load with each cycle is $\Delta \langle n \rangle_L=1/2$. 

To generalize, for the specified initial condition we can write the  probability of occupation, $p_{n}^{\mathcal{C}_{N_c}}$, of the $n-$th level of the harmonic oscillator.  The superscript ~$\mathcal{C}_{N_c}$ denotes that this is the probability of occupation after stroke (III) (which brings the ion to point $\mathcal{C}$ in Fig. 1(c)), of the $N_c-$th cycle. We get
\begin{align}
p_{n}^{\mathcal{C}_{N_c}} &= \sum_{k=0}^{N_c} \mathrm{C}^{N_c}_{k} \; (p^{\mathcal{A}}_D)^{k} (1-p^{\mathcal{A}}_D)^{N_c-k} \delta_{n,2k+{\rm mod}(N_c,2)} \label{eq:biased_diff}
\end{align}
where $\mathrm{C}^b_a=b!/[a! (b-a)!]$, once again $p^{\mathcal{A}}_D$ is the value to which the engine population is reset once in each cycle, at the end of stroke (I), and $\delta_{a,b}$ is the Kronecker delta function. The sum over integers $k$, together with the Kronecker delta, ensures that only even (odd) levels are occupied for even (odd) values of $N_c$. Eq.(\ref{eq:biased_diff}) describes the evolution of a probability distribution of a biased normal diffusive process, and this implies three things: (i) that the change in average phonon occupation of the load is $\Delta \langle n \rangle_L =1/2$ with each additional cycle, (ii) that the entropy of the load grows asymptotically as the logarithm of the number of cycles, i.e. $\mathcal{S}_L \propto ln{N_c}$~, and (iii) that the distribution will tend towards a Gaussian. This last point is particularly important because it implies that the load is not in a passive state, and that its ergotropy increases with $N_c$. From Eq.(\ref{eq:biased_diff}) we can compute that the ergotropy grows as $\mathcal{E}_L\propto N_c-\alpha\sqrt{N_c}$, where $\alpha$ is a constant. This expression comes from the fact that the energy of the load grows linearly with $N_c$, while the energy of the ``passified'' load only grows as $\sqrt{N_c}$.   
\end{appendix}

\end{document}